\documentclass[english,conference,10pt]{IEEEtran}
\usepackage[T1]{fontenc}
\usepackage[latin9]{inputenc}
\usepackage{amsmath}
\usepackage{amssymb}
\usepackage{stmaryrd}
\usepackage{graphicx}
\usepackage{esint}

\makeatletter

\usepackage{amsmath}
\usepackage{mathabx}

\DeclareMathOperator{\enum}{e}
\DeclareMathOperator{\inum}{i}

\DeclareMathOperator{\sinc}{sinc}

\DeclareMathOperator{\erfc}{erfc}
\DeclareMathOperator{\BER}{BER}

\allowdisplaybreaks
\IEEEoverridecommandlockouts

\author{
\IEEEauthorblockN{Sander Wahls\IEEEauthorrefmark{1},
Son T. Le\IEEEauthorrefmark{2},
Jaroslaw E. Prilepsky\IEEEauthorrefmark{2},
H. Vincent Poor\IEEEauthorrefmark{3} and
Sergei K. Turitsyn\IEEEauthorrefmark{2}
}

\vspace{0.1in}

\IEEEauthorblockA{\IEEEauthorrefmark{1}Delft Center for Systems and Control, TU Delft, The Netherlands. Email: s.wahls@tudelft.nl}

\IEEEauthorblockA{\IEEEauthorrefmark{2}Aston Institute of Photonic Technologies, Aston University, UK. Email: \{let1,y.prylepskiy1,s.k.turitsyn\}@aston.ac.uk}

\IEEEauthorblockA{\IEEEauthorrefmark{3}Department of Electrical Engineering, Princeton University, USA. Email: poor@princeton.edu}
}

\makeatother

\usepackage{babel}
\begin{document}

\title{Digital Backpropagation in the\\
Nonlinear Fourier Domain\thanks{This research was supported in parts by the UK EPSRC programme grant UNLOC (EP/J017582/1) and  by the U. S. National Science Foundation under Grant CCF-1420575.}}
\maketitle
\begin{abstract}
Nonlinear and dispersive transmission impairments in coherent fiber-optic
communication systems are often compensated by reverting the nonlinear
Schr\"odinger equation, which describes the evolution of the signal
in the link, numerically. This technique is known as digital backpropagation.
Typical digital backpropagation algorithms are based on split-step
Fourier methods in which the signal has to be discretized in time
and space. The need to discretize in both time and space however makes
the real-time implementation of digital backpropagation a challenging
problem. In this paper, a new fast algorithm for digital backpropagation
based on nonlinear Fourier transforms is presented. Aiming at a proof
of concept, the main emphasis will be put on fibers with normal dispersion
in order to avoid the issue of solitonic components in the signal.
However, it is demonstrated that the algorithm also works for anomalous
dispersion if the signal power is low enough. Since the spatial evolution
of a signal governed by the nonlinear Schr\"odinger equation can
be reverted analytically in the nonlinear Fourier domain through simple
phase-shifts, there is no need to discretize the spatial domain. The
proposed algorithm requires only $\mathcal{O}(D\log^{2}D)$ floating
point operations to backpropagate a signal given by $D$ samples,
independently of the fiber's length, and is therefore highly promising
for real-time implementations. The merits of this new approach are
illustrated through numerical simulations.
\end{abstract}


\section{Introduction}

The evolution of a signal $E(x,t)\in\mathbb{C}$ in optical fiber,
where $x\ge0$ denotes the position in the fiber and $t\in\mathbb{R}$
the time, is well-described through the \emph{nonlinear Schr\"odinger
equation (NSE)}. Through proper scaling and coordinate transforms,
the NSE can be brought into its normalized form
\begin{equation}
\inum\frac{\partial E}{\partial x}+\frac{\partial^{2}E}{\partial t^{2}}+2\kappa|E|^{2}E=-\inum\Gamma E,\quad\kappa\in\{\pm1\}.\label{eq:NSE}
\end{equation}
The parameter $\kappa$ in Eq. (\ref{eq:NSE}) effectively determines
whether the fiber dispersion is normal ($-1$) or anomalous ($+1$),
while the parameter $\Gamma$ determines the loss in the fiber. In
the following, it will be assumed that the loss parameter is zero.
There are two important cases in which a fiber-optic communication
channel can be modeled under this assumption. When the fiber loss
is mitigated through periodic amplification of the signal, the average
of the properly transformed signal is known to satisfy the NSE with
zero loss \cite{Hasegawa1991b}. Furthermore, recently a distributed
amplification scheme with an effectively unattenuated optical signal
(quasi-lossless transmission directly described by the lossless NSE)
has been demonstrated \cite{Ania-Castanon2008}. If the loss parameter
is zero, the NSE can be solved using \emph{nonlinear Fourier transforms
(NFTs)} \cite{Zakharov1972}. The spatial evolution of the signal
$E(x,t)$ then reduces to a simple phase-shift in the \emph{nonlinear
Fourier domain (NFD)}, similar to how linear convolutions reduce to
phase-shifts in the conventional Fourier domain. The prospect of an
optical communication scheme that inherently copes with the nonlinearity
of the fiber has recently led to several investigations on how data
can be transmitted in the NFD instead of the conventional Fourier
or time domains \cite{Yousefi2014compact,Turitsyna2013,Prilepsky2013,Prilepsky2014a,Le2014,Hari2014,Zhang2014,Buelow2015},
with the original idea being due to Hasegawa and Nyu \cite{Hasegawa1993}.
Next to potential savings in computational complexity, it is anticipated
that subchannels defined in the NFD will not suffer from intra-channel
interference, which is currently limiting the data rates achievable
by wavelength-division multiplexing systems \cite{Essiambre2010}. 

\begin{figure}
\centering{}\includegraphics[width=1\columnwidth]{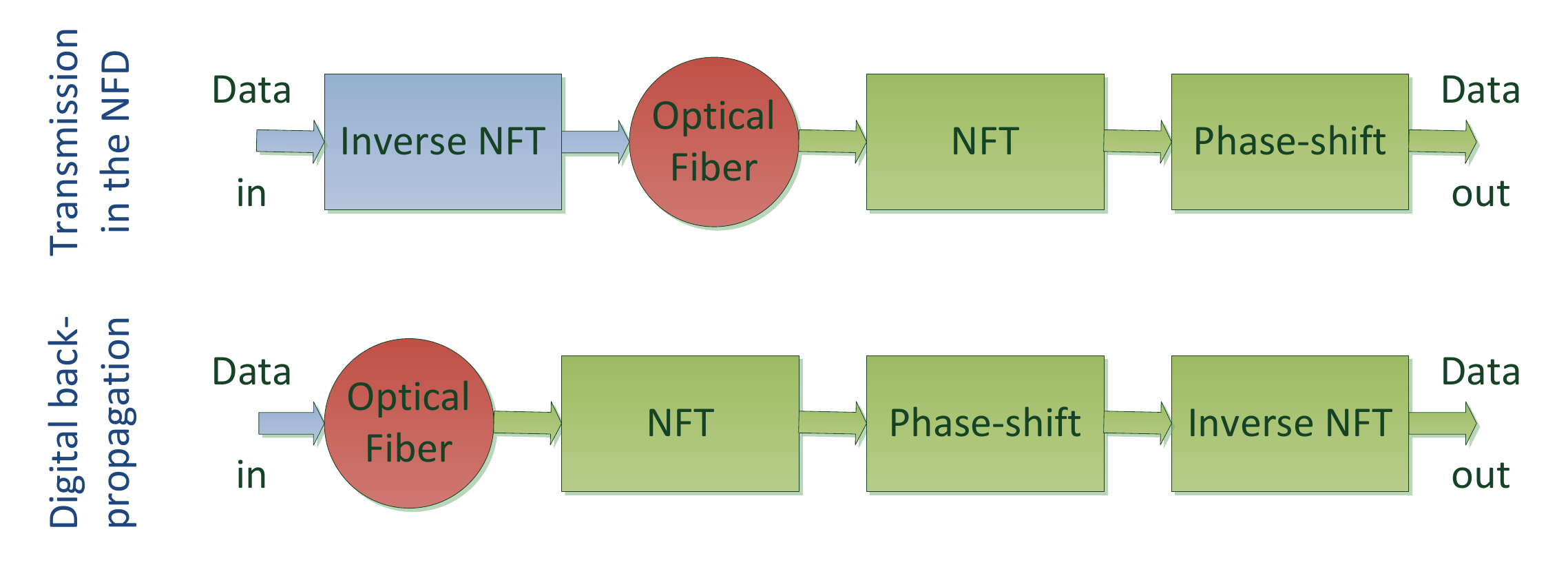}\caption{\label{fig:mod-vs-db}Information transmission (top) vs DBP (bottom)
in the NFD}
\end{figure}

The mathematics behind the NFT are however quite involved, and despite
recent progress in the implementation of fast forward and inverse
NFTs \cite{Wahls2013b,Wahls2013d,Wahls2015b} no integrated concept
for a computationally efficient fiber-optic transmission system that
operates in the NFD seems to be available. In this paper, therefore
the problem of \emph{digital backpropagation (DBP)}, i.e. recovering
the fiber input $E(0,t)$ from the output $E(x_{1},t)$ by solving
(\ref{eq:NSE}), is addressed instead using NFTs \cite{Turitsyna2013}.
Both concepts are compared in Fig. \ref{fig:mod-vs-db}. Although
DBP does not solve the issue of intra-channel interference in fiber-optic
networks because it is usually not feasible to join the individual
subchannels of the physically separated users into a single super-channel
in this case \cite[X.D]{Essiambre2010}, we note that there are other
scenarios where this issue does not arise \cite{Maher2015}. The advantage
of digital backpropagation over information transmission in the NFD
is that it is not necessary to implement full forward and backward
NFTs. Together with recent advantages made in \cite{Wahls2013b,Wahls2013d,Wahls2015b},
this observation will enable us to perform digital backpropagation
in the NFD using only $\mathcal{O}(D\log^{2}D)$ floating point operations
(\emph{flops}), where $D$ is the number of samples. This complexity
estimate is independent of the length of the fiber because the spatial
evolution of the signal will be carried out analytically. Conventional
split-step Fourier methods, in contrast, have a complexity of $\mathcal{O}(MD\log D)$
flops, where $M$ is the number of spatial steps \cite[III.G]{Ip2008}.

The goal of this paper is to present a new, fast algorithm for digital
backpropagation that operates in the NFD and to compare it with traditional
split-step Fourier methods through numerical simulations. The impact
of noise resulting from the use of distributed Raman amplification
will be of particular interest. The paper is structured as follows.
In Sec. \ref{sec:DB-in-NFD}, the theory behind digital backpropagation
in the NFD will be outlined, and the new, fast algorithm will be given
and discussed. The simulation setup is described in Sec. \ref{sec:Simulation-Setup},
while results are reported in Sec. \ref{sec:Simulation-Results}.
Sec. \ref{sec:Conclusion} concludes the paper.

\section{Digital Backpropagation in the\protect \\
Nonlinear Fourier Domain\label{sec:DB-in-NFD}}

In this section, first the theoretical and computational results that
are required to perform digital backpropagation in the NFD are briefly
recapitulated from \cite{Wahls2015b}. Afterwards, the fast algorithm
is presented and its limitations are discussed.

\subsection{Theory for the Continuous-Time Case}

The \emph{Zakharov-Shabat scattering problem} associated to any signal
$E(x,t)$ that vanishes sufficiently fast for $t\to\pm\infty$ is
\begin{align}
\frac{d}{dt}\boldsymbol{\mathbf{\phi}}(x,t,\lambda)= & \left[\begin{array}{cc}
-\inum\lambda & E(x,t)\\
-\kappa\bar{E}(x,t) & \inum\lambda
\end{array}\right]\boldsymbol{\mathbf{\phi}}(x,t,\lambda),\label{eq:Z-S-1}\\
\boldsymbol{\phi}(x,t,\lambda)= & \left[\begin{array}{c}
\enum^{-\inum\lambda t}\\
0
\end{array}\right]+o(1),\quad t\to-\infty.\label{eq:Z-S-2}
\end{align}
With $\phi_{1}$ and $\phi_{2}$ denoting the components of $\boldsymbol{\phi}$,
now define
\begin{align}
\alpha(x,\lambda):= & \lim_{t\to\infty}\enum^{\inum\lambda t}\phi_{1}(x,t,\lambda),\label{eq:alpha(x,lambda)}\\
\beta(x,\lambda):= & \lim_{t\to\infty}\enum^{-\inum\lambda t}\phi_{2}(x,t,\lambda),\nonumber 
\end{align}
where $\lambda\in\mathbb{C}$ is a parameter. If $E(x,t)$ satisfies
the NSE (\ref{eq:NSE}) with zero loss parameter $\Gamma=0$, the
corresponding $\alpha(x,\lambda)$ and $\beta(x,\lambda)$ turn out
to depend on $x$ in very simple way: 
\begin{equation}
\alpha(x,\lambda)=\alpha(0,\lambda),\quad\beta(x,\lambda)=\enum^{-4\inum\lambda^{2}x}\beta(0,\lambda).\label{eq:time-evolution-al-be}
\end{equation}
These functions are not the final form of the NFT, but for our needs
it will be sufficient to stop here.

The complicated spatial evolution of the signal $E(x,t)$ thus indeed
becomes trivial if it is transformed into $\alpha(x,\lambda)$ and
$\beta(x,\lambda)$. Based on this insight, ideal continuous-time
digital backpropagation reduces to three basic steps in the NFD:
\begin{equation}
E(x_{1},t)\stackrel{\text{A}}{\to}\left[\begin{array}{c}
\alpha(x_{1},\lambda)\\
\beta(x_{1},\lambda)
\end{array}\right]\stackrel{\text{B}}{\to}\left[\begin{array}{c}
\alpha(0,\lambda)\\
\beta(0,\lambda)
\end{array}\right]\stackrel{\text{C}}{\to}E(0,t).\label{eq:three-steps}
\end{equation}

\subsection{Discretization of the Continuous-Time Problem}

In order to obtain a numerical approximation of $\alpha(x,\lambda)$
and $\beta(x,\lambda)$ for any fixed $x$, choose a sufficiently
large interval $[T_{1},T_{2}]$ inside which $E(x,t)$ has already
vanished sufficiently. Without loss of generality, one can assume
that $T_{1}=-1$ and $T_{2}=0$. In this interval, $D$ rescaled samples
\[
E[x,n]:=\epsilon E\left(x,-1+n\epsilon-\frac{\epsilon}{2}\right),\,\epsilon:=\frac{1}{D},\,n\in\{1,\dots,D\},
\]
are taken. With $z:=\enum^{-2\inum\lambda\epsilon}$, (\ref{eq:Z-S-1})--(\ref{eq:Z-S-2})
becomes 
\begin{align}
\boldsymbol{\phi}[x,n,z]:= & z^{\frac{1}{2}}\left[\begin{array}{cc}
1 & z^{-1}E[x,n]\\
-\kappa\bar{E}[x,n] & z^{-1}
\end{array}\right]\nonumber \\
 & \times\frac{\boldsymbol{\phi}[x,n-1,z]}{\sqrt{1+\kappa|E[x,n]|^{2}}},\label{eq:discrete-Z-S-1}\\
\boldsymbol{\phi}[x,0,z]:= & z^{-\frac{D}{2}}\left[\begin{array}{c}
1\\
0
\end{array}\right].\label{eq:discrete-Z-S-2}
\end{align}
This leads to the following polynomial approximations: 
\begin{align*}
\alpha(x,\lambda)\approx a(x,z):= & {\textstyle \sum_{i=0}^{D-1}}a_{i}(x)z^{-i}:=\phi_{1}[x,D,z],\\
\beta(x,\lambda)\approx b(x,z):= & {\textstyle \sum_{i=0}^{D-1}}b_{i}(x)z^{-i}:=\phi_{2}[x,D,z].
\end{align*}

\subsection{The Algorithm}

The three steps in the diagram (\ref{eq:three-steps}) can be implemented
with an overall complexity of $\mathcal{O}(D\log^{2}D)$ flops as
follows.

\subsubsection*{Step A }

The discrete scattering problem to find the polynomials $a(x_{1},z)$
and $b(x_{1},z)$ from the known fiber output $E(x_{1},t)$ through
(\ref{eq:discrete-Z-S-1})--(\ref{eq:discrete-Z-S-2}) is solved with
only $\mathcal{O}(D\log^{2}D)$ flops using \cite[Alg. 1]{Wahls2013d}
(also see \cite[Alg. 1]{Wahls2013b}).

\subsubsection*{Step B}

The unknown polynomials $a(0,z)$ and $b(0,z)$ are defined by the
unknown fiber input $E(0,t)$ through (\ref{eq:discrete-Z-S-1})--(\ref{eq:discrete-Z-S-2}).
In this step, approximations $\hat{a}(0,z)$ and $\hat{b}(0,z)$ of
$a(0,z)$ and $b(0,z)$ are computed based on Eq. (\ref{eq:time-evolution-al-be}).
The left-hand in (\ref{eq:time-evolution-al-be}) suggests the choice
$\hat{a}(0,z):=\hat{a}(x_{1},z)$. For finding $\hat{b}(0,t)$, let
us denote the $D$-th root of unity by $w:=\enum^{-2\pi\inum/D}$.
The right-hand in (\ref{eq:time-evolution-al-be}) motivates us to
find $\hat{b}(0,z)=\sum_{i=0}^{D-1}\hat{b}_{i}(0)z^{-i}$ by solving
the well-posed interpolation problem
\begin{align}
\hat{b}(0,w^{n-\frac{1}{2}})= & \enum^{4\inum\left(\log(w^{n-\frac{1}{2}})/(-2\inum\epsilon)\right)^{2}x_{1}}b(x_{1},w^{n-\frac{1}{2}})\label{eq:interpolation-problem}\\
= & \enum^{4\pi^{2}\inum(n-\frac{1}{2})^{2}x_{1}}b(x_{1},w^{n-\frac{1}{2}}),\,n\in\{1,\dots,D\},\nonumber 
\end{align}
with the fast Fourier transform, using only $\mathcal{O}(D\log D)$
flops.

\subsubsection*{Step C}

The inverse scattering problem of estimating $E[0,n]$ from $\hat{a}(0,z)$
and $\hat{b}(0,z)$ by inverting (\ref{eq:discrete-Z-S-1})--(\ref{eq:discrete-Z-S-2})
is solved using $\mathcal{O}(D\log^{2}D)$ flops as described in \cite[IV]{Wahls2015b}.

\subsection{Limitations\label{sub:Limitations}}

It was already mentioned above that the algorithm works for fibers
with normal dispersion. In that case, the sign $\kappa$ in the NSE
(\ref{eq:NSE}) will be negative and $E(x,t)$ is determined through
the values that $\alpha(0,\lambda)$ and $\beta(0,\lambda)$ take
on the real axis $\mathbb{R}\ni\lambda$ \cite[p. 285]{Ablowitz1974}.
If the dispersion is anomalous, the sign will be positive and $E(x,t)$
is determined through the values that $\alpha(0,\lambda)$ and $\beta(0,\lambda)$
take on the real axis \emph{and} around the roots of $\alpha(0,\lambda)$
in the complex upper half-plane $\Im(\lambda)>0$ \cite[IV.B]{Ablowitz1974}.
Taking the coordinate transform $z=\enum^{-2\inum\lambda\epsilon}$
into account, one sees that the interpolation problem (\ref{eq:interpolation-problem})
however enforces the phase-shift only for certain real values of $\lambda$.
Thus, the algorithm is unlikely to work if $\alpha(0,\lambda)$ has
roots in the upper half-plane. The condition $\int_{-\infty}^{\infty}|E(0,t)|dt<\frac{\pi}{2}$
is sufficient to ensure that there are no such roots \cite[Th. 4.2]{Klaus2003}.
The actual threshold where solitons start to emerge however is expected
to be higher due to the randomly oscillating character of waveforms
used in fiber-optic communications \cite{Turitsyn2008,Derevyanko2008}.

\section{Simulation Setup\label{sec:Simulation-Setup}}

\begin{figure}

\begin{centering}
\includegraphics[width=1\columnwidth]{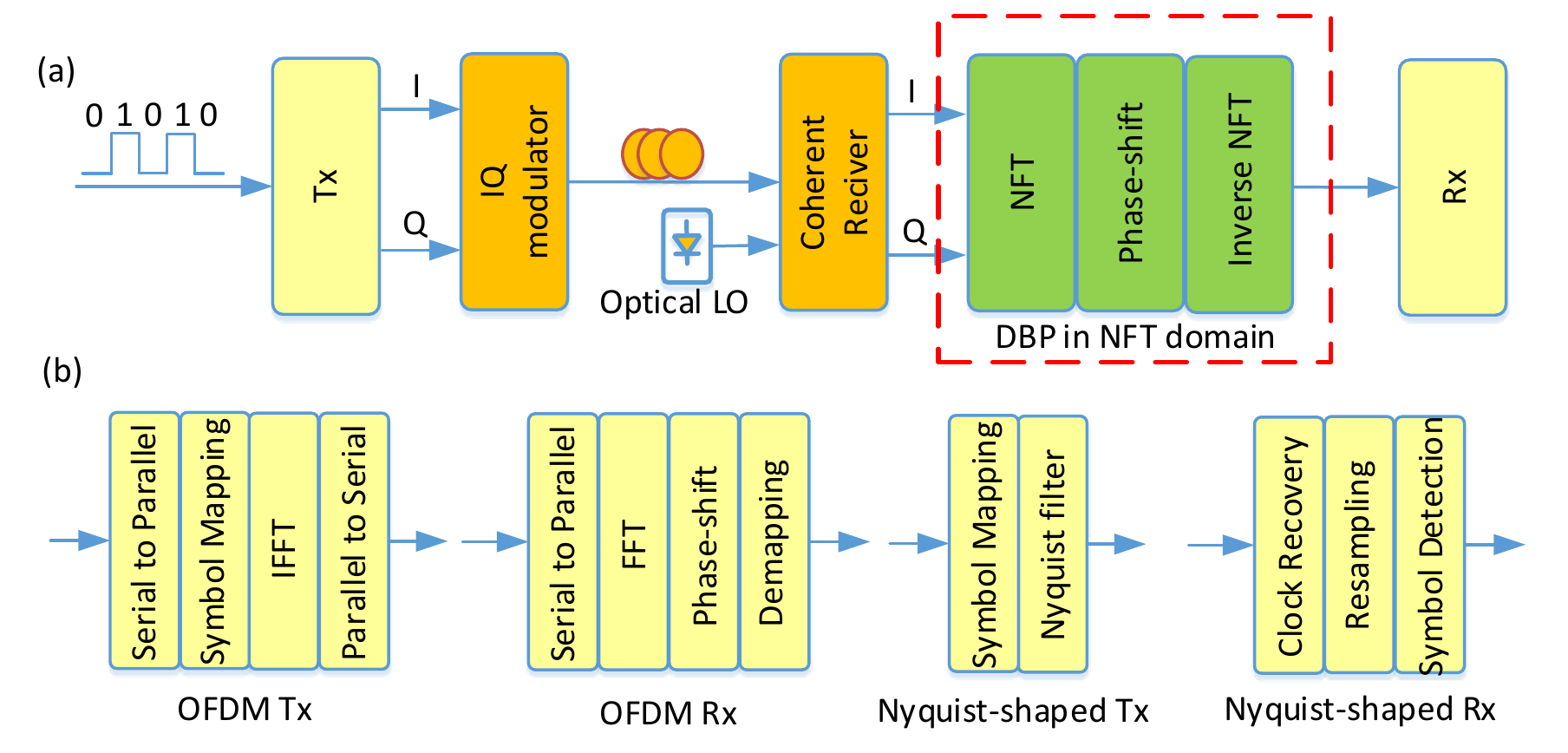}
\par\end{centering}

\caption{\label{fig:Simulation-setup}a) Simulation setup of coherent optical
communication systems with DBP in the NFD, b) basic block functions
of OFDM and Nyquist transceivers}

\end{figure}

In this section, the simulation setup that was used to assess the
performance of DBP in the NFD is presented.

The transmission link was assumed to be lossless due to ideal Raman
amplification, with an \emph{amplified spontaneous emission (ASE)}
noise density $N_{\text{ASE}}=\Gamma Lhf_{s}K_{T}$. Here, $\Gamma$
is the fiber loss, $L$ is the transmission distance, $hf_{s}$ is
the photon energy, $f_{s}$ is the optical frequency of the Raman
pump providing the distributed gain, and $K_{T}=1.13$ is the photon
occupancy factor for Raman amplification of a fiber-optic communication
system at room temperature. In the simulations, it was assumed that
the long-haul fiber link consisted of $80$-km spans of fiber (\emph{standard
single-mode fiber (SSMF)} in the anomalous case) with a loss of $0.2$
dB/km, nonlinearity coefficient of $1.22$ W/km, and a dispersion
of $\pm16$ ps/nm/km (normal and anomalous dispersions). A photon
occupancy factor of $4$ was used for more realistic conditions. The
ASE noise was added after each fiber span.

The data was modulated using high spectral efficiency modulation formats
(QPSK and 64QAM) and either \emph{Nyquist pulses} (i.e. $\sinc$'s)
\cite{Le2014} or \emph{orthogonal frequency division multiplexing
(OFDM)} \cite{Prilepsky2013,Prilepsky2014a,Le2014}. The block diagram
of the simulation setup and basic block functions of the OFDM and
Nyquist transceivers are presented in Fig. \ref{fig:Simulation-setup}.
In Fig. \ref{fig:Simulation-setup}(a), DBP in the NFD is performed
at the receiver after coherent detection, synchronization, windowing
and frequency offset compensation. For simplicity, both perfect synchronization
and frequency offset compensation were assumed. The net data rates
of the considered transmission systems were, after removing $7\%$
overhead due to the \emph{forward error correction (FEC)}, $100$
Gb/s and $300$ Gb/s for QPSK and 64QAM, respectively. For the OFDM
system, the size of the \emph{inverse fast Fourier transform (IFFT)}
was $128$ samples, and $112$ subcarriers were filled with data using
\emph{Gray coding}. The remaining subcarriers were set to zero. The
useful OFDM symbol duration was $2$ ns and no cyclic prefix was used
for the linear dispersion removal. An oversampling factor of $8$
was adopted resulting in a total simulation bandwidth of \textasciitilde{}$448$GHz.
At the receiver side, all digital signal processing operations were
performed with the same sampling rate. The receiver\textquoteright s
bandwidth is assumed to be unlimited in order to estimate the achievable
gain offered by the proposed DBP algorithm. 

\begin{figure}
\begin{centering}
\includegraphics[width=0.9\columnwidth]{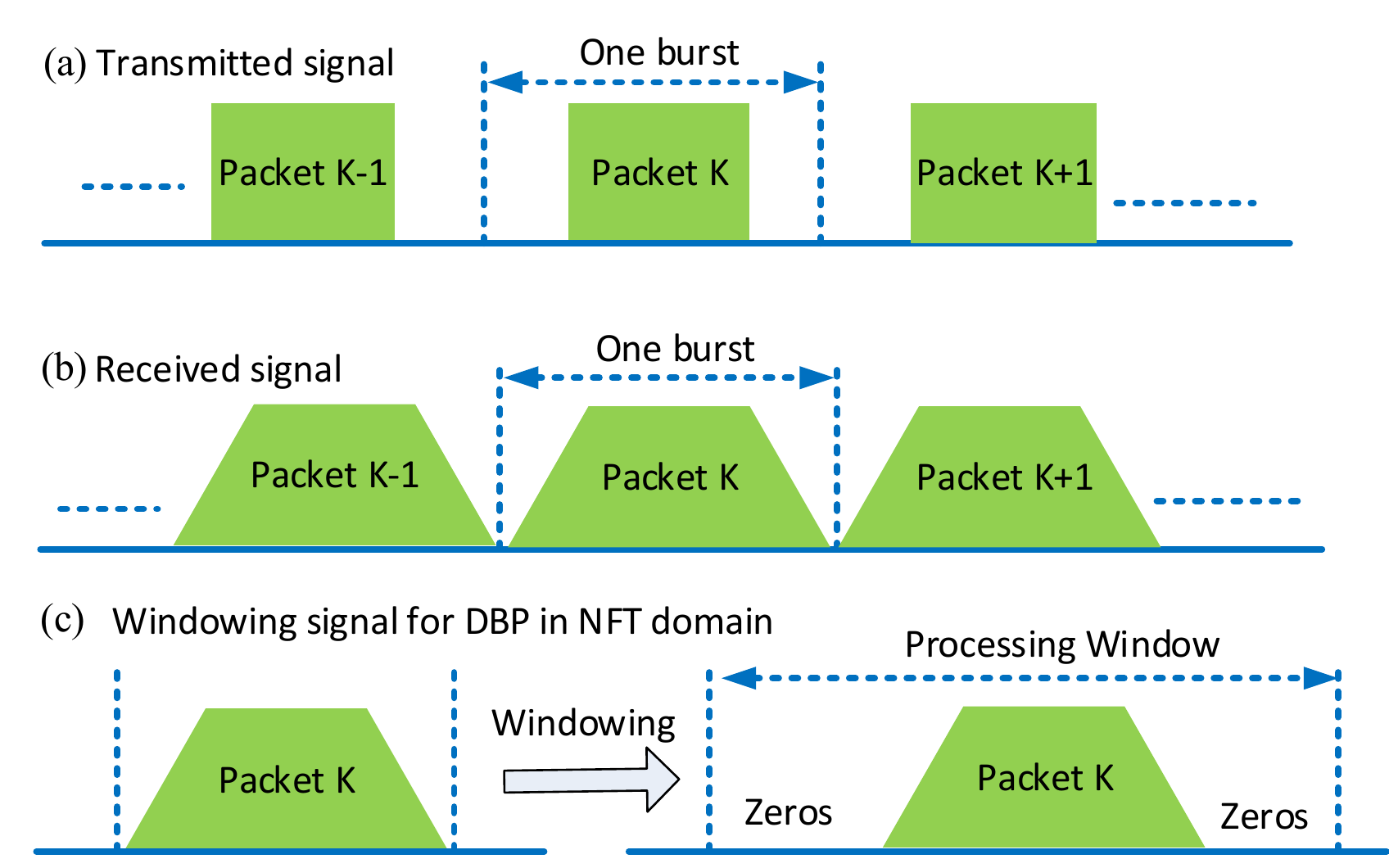}
\par\end{centering}

\caption{\label{fig:burst-mode}a) Illustration of a burst mode transmission
at the transmitter side, in which neighboring packets are separated
by a guard time, b) received signal at the receiver, which is broadened
by chromatic dispersion, c) windowing signal for processing in the
proposed nonlinear compensation scheme.}
\end{figure}

The data was transmitted in a burst mode where data packets where
separated by a guard time. See Fig. \ref{fig:burst-mode} for an illustration.
The guard time was chosen longer than the memory $\Delta T=2\pi B\beta_{2}L$
induced by the fiber \emph{chromatic dispersion (CD)}, where \textbf{$B$}
is the signal\textquoteright s bandwidth, $\beta_{2}$ is the chromatic
dispersion and $L$ is the transmission distance. One burst consisted
of one data packet and the associated guard interval. At the receiver,
after synchronization, each burst is extracted and processed separately.
Since the forward and inverse NFTs require that the signal has vanished
early enough before it reaches the boundaries, zero padding was applied
to enlarge the processing window as in Fig. \ref{fig:burst-mode}(c).

\section{Simulation Results\label{sec:Simulation-Results}}

In this section, the performance of DBP in the NFD is compared with
a traditional DBP algorithm based on the split-step Fourier method
\cite{Ip2008} as well as with simple chromatic dispersion compensation;
the latter works well only in the low-power regime where nonlinear
effects are negligible.

\subsection{Normal dispersion}

In the normal dispersion case, a $56$ Gbaud Nyquist-shaped transmission
scheme is considered in burst mode with $256$ symbols in each packet.
The duration of each packet is \textasciitilde{}$4.6$ns. The burst
size is $16000$ samples and the processing window size for each burst
after zero padding is $D=65536$ samples. The guard time is \textasciitilde{}$10\%$
longer than the fiber chromatic dispersion induced memory of the link.
The forward propagation is simulated using the split-step Fourier
method \cite{Ip2008} with $80$ steps/span, i.e. $1$km step. Monte-Carlo
simulations were performed to estimate the system performance using
the \emph{error vector magnitude (EVM)} \cite[(5)]{Shafik2006}. For
convenience, the EVM is then converted into the \emph{$Q$-factor}
$20\log(\sqrt{2}\erfc^{-1}(2\BER))$ using the \emph{bit error rate
(BER)} estimate \cite[(13)]{Shafik2006}. 

\begin{figure}
\begin{centering}
\includegraphics[width=0.9\columnwidth]{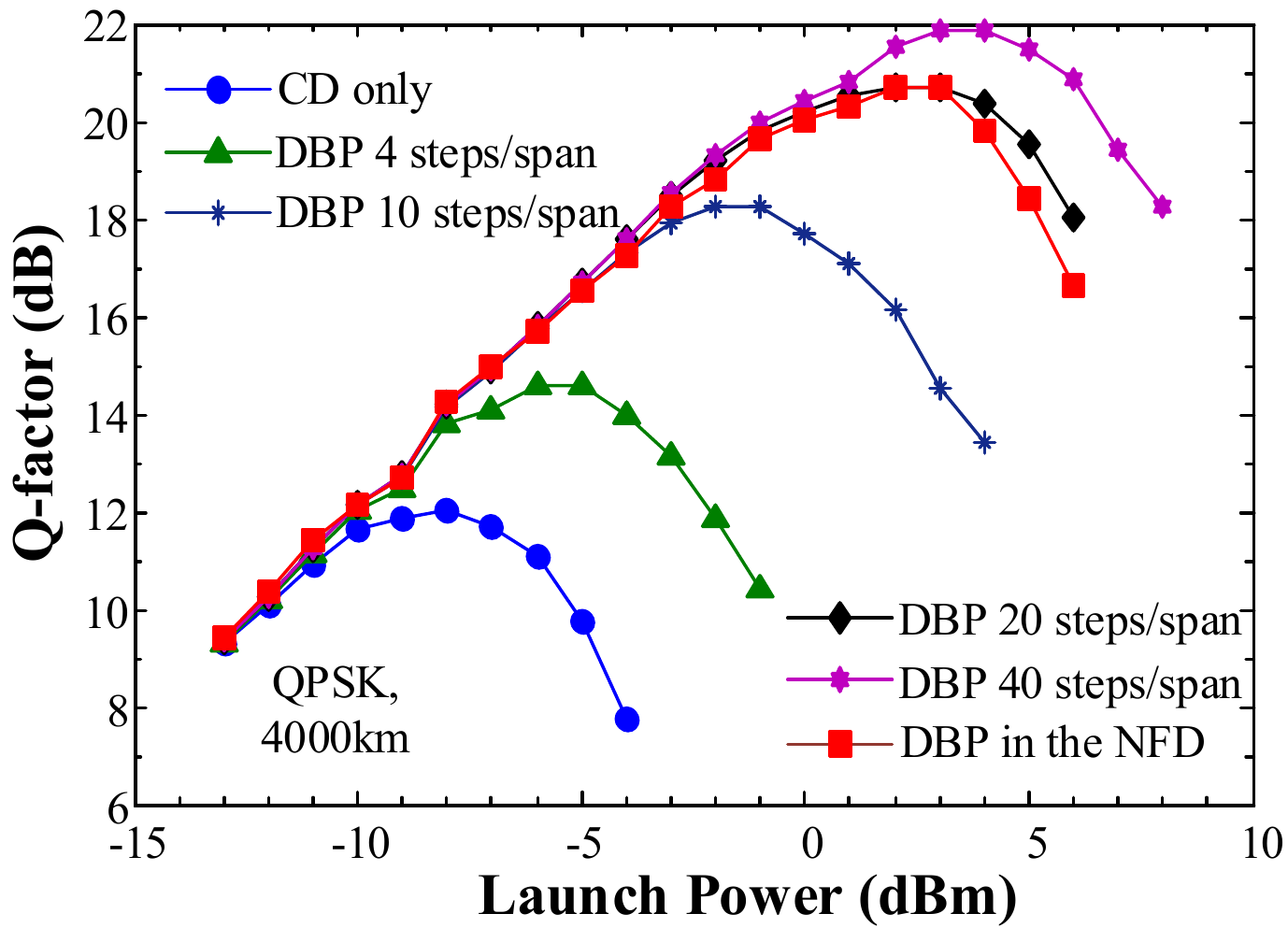}
\par\end{centering}

\caption{\label{fig:Simulation-results-1} Performance of 100-Gb/s QPSK Nyquist-shaping
over 4000km.}
\end{figure}

\begin{figure}
\begin{centering}
\includegraphics[width=0.9\columnwidth]{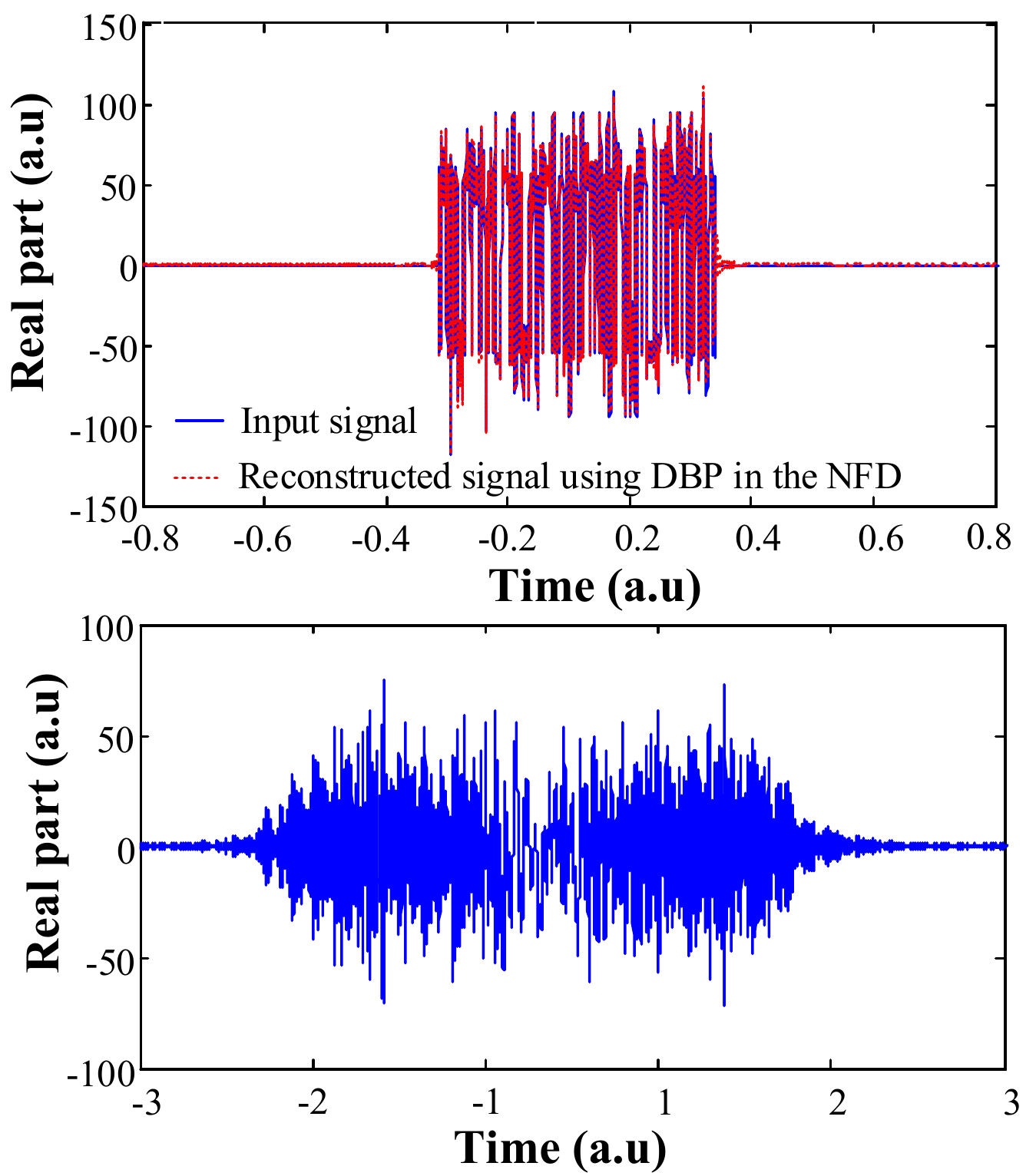}
\par\end{centering}

\caption{\label{fig:Simulation-results-0} Top: True vs reconstructed (via
DBP-NFD) fiber input for a QPSK-Nyquist signal. Bottom: Corresponding
fiber output. Only real parts are shown.}
\end{figure}

The performance of the $100$-Gb/s QPSK Nyquist-shaped system is depicted
in Fig. \ref{fig:Simulation-results-1} as a function of the launch
power in a 4000km link for various configurations. An exemplary fiber
in- and output as well as the corresponding reconstructed input (via
DBP in the NFD) are shown in Fig. \ref{fig:Simulation-results-0}.
It can be seen in Fig. \ref{fig:Simulation-results-1} that the proposed
DBP in the NFD provides a significant performance gain of \textasciitilde{}$8.6$
dB, which is comparable with the traditional DBP employing $20$ steps/span.
Traditional DBP with 40 steps/span can be considered as ideal DBP
in this experiment because a further increase of the number of steps/span
did not improve the performance further. In the considered $4000$km
link, $40$ steps/span DBP requires $M=2000$ steps in total. This
illustrates the advantage of the proposed DBP algorithm whose complexity
is in contrast independent of the transmission distance. The performance
of DBP in the NFD was however found to degrade rapidly when the launch
power is sufficiently high. We believe that this effect can be mitigated
through an extension of the processing window at the cost of increased
computational complexity.


\begin{figure}
\begin{centering}
\includegraphics[width=0.9\columnwidth]{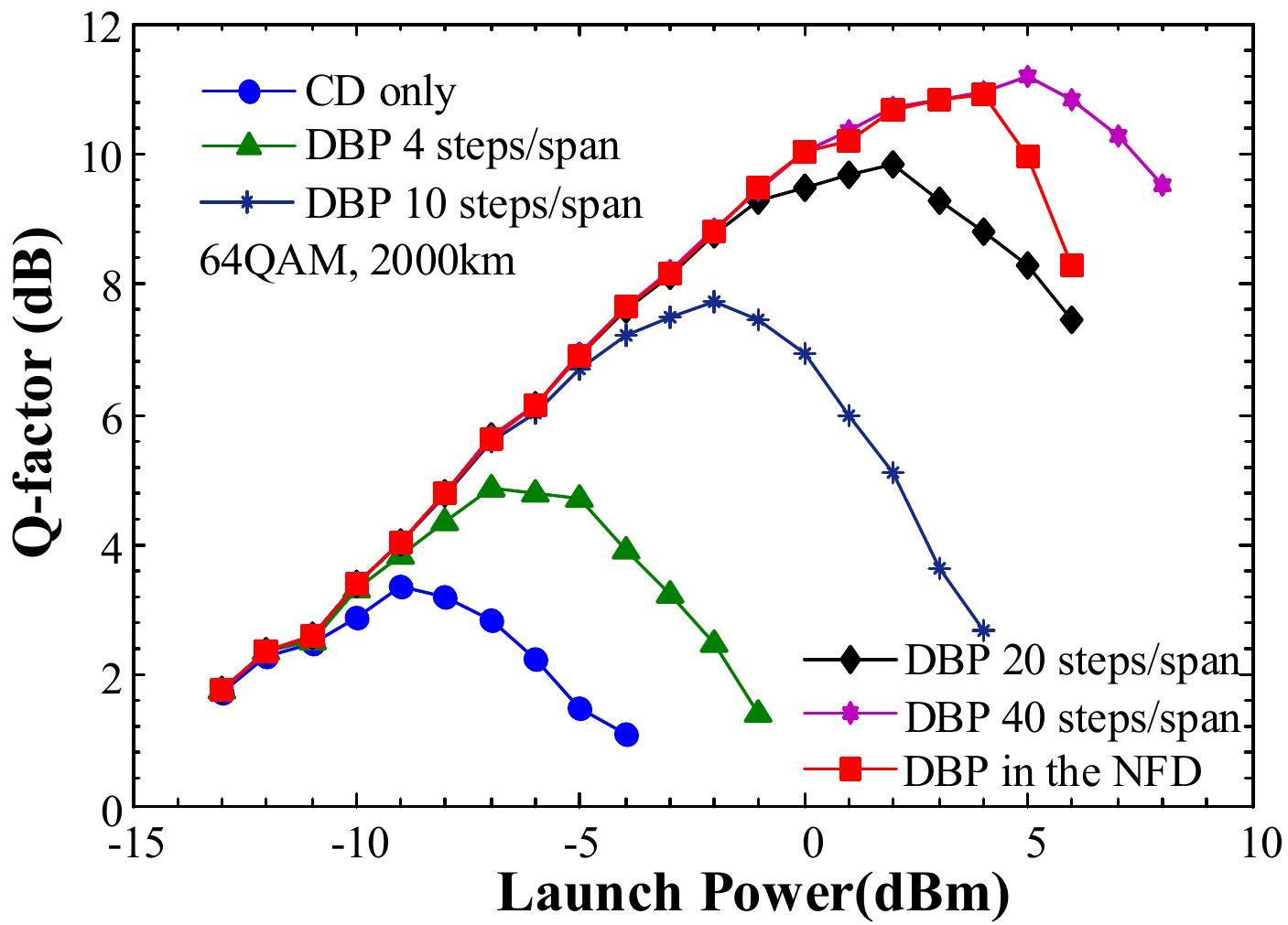}
\par\end{centering}

\caption{\label{fig:Simulation-results-2} Performance of 300-Gb/s 64QAM Nyquit-shaping
over 2000km.}
\end{figure}

A similar behavior can be observed in Fig. \ref{fig:Simulation-results-2}
for a $300$-Gb/s 64QAM Nyquist-shaped system. It can be seen in Fig.
\ref{fig:Simulation-results-2} that DBP in the NFD shows \textasciitilde{}$1.5$
dB better performance than DBP employing $20$ steps/span. The difference
from ideal DBP ($40$ steps/span) is just \textasciitilde{}$0.6$
dB. This clearly shows that NFD-DBP can provide essentially the same
performance as ideal DBP. 

\subsection{Anomalous dispersion}

Most fibers used today (such as SSMF) have anomalous dispersion. Therefore,
it is critical to evaluate the performance of the proposed DBP algorithm
in optical links with anomalous dispersion. The DBP-NFD algorithm
proposed earlier however cannot yet deal with signals where the function
$\alpha(0,\lambda)$ defined in (\ref{eq:alpha(x,lambda)}) has roots
in the upper half-plane. See Sec. \ref{sub:Limitations}. The condition
that the $L_{1}$-norm of each burst is less than $\pi/2$ is sufficient
to ensure this condition. The packet duration in a system with anomalous
dispersion thus has to be kept small enough in order to apply the
proposed algorithm. This constraint is not desirable in practice,
as it reduces the total throughput of the link because the guard interval,
which is independent of the packet duration, must be inserted more
frequently. The $L_{1}$ norm of the Fourier transform is always lower
or equal to the $L_{1}$ norm of the time-domain signal. Motivated
by this observation, OFDM was chosen instead of Nyquist-shaping for
anomalous dispersion. 

\begin{figure}
\begin{centering}
\includegraphics[width=0.9\columnwidth]{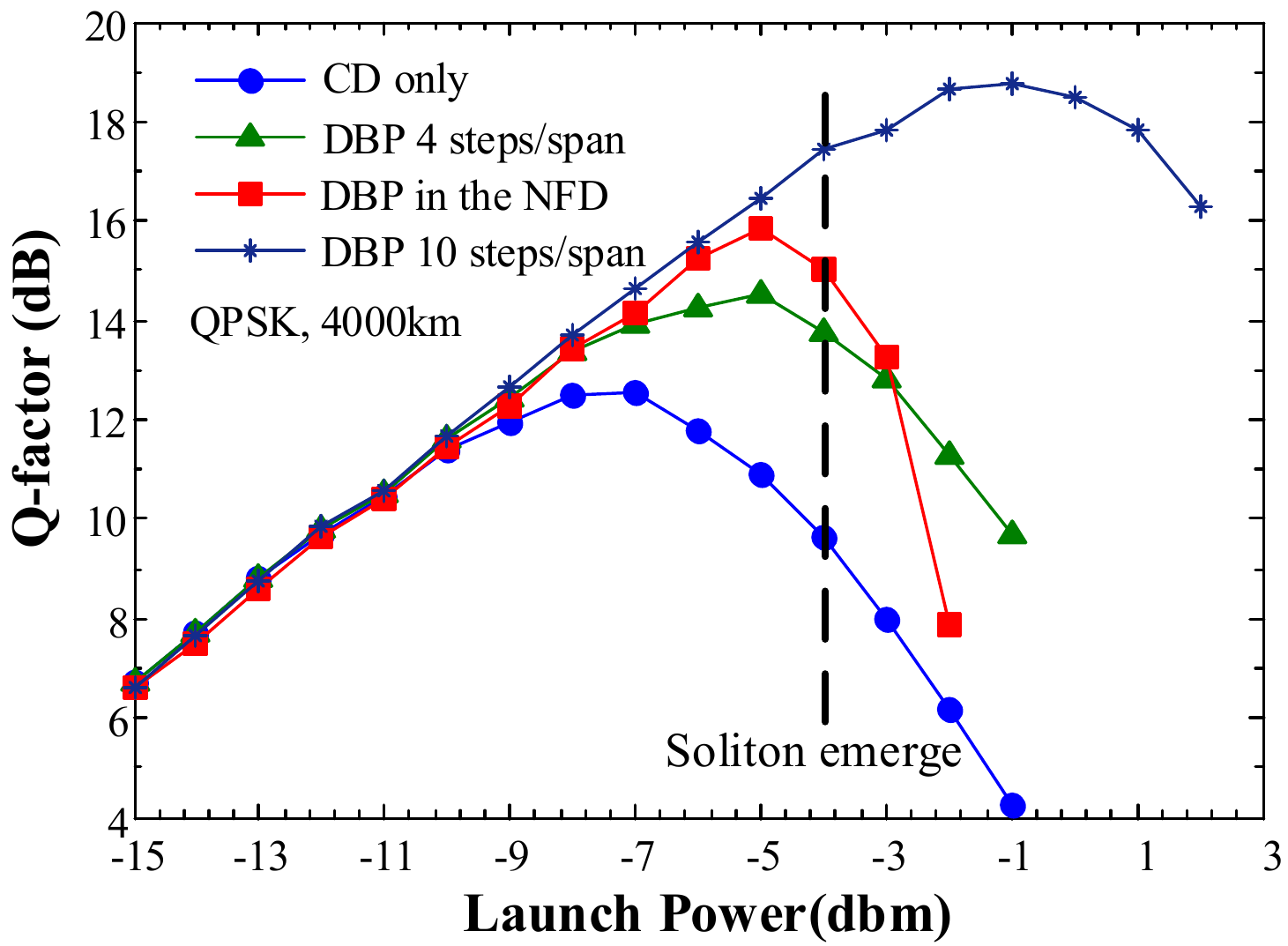}
\par\end{centering}

\caption{\label{fig:Simulation-results-3} Performance of 100-Gb/s QPSK OFDM
over 4000km in fiber with anomalous dispersion.}
\end{figure}

\begin{figure}
\begin{centering}
\includegraphics[width=0.9\columnwidth]{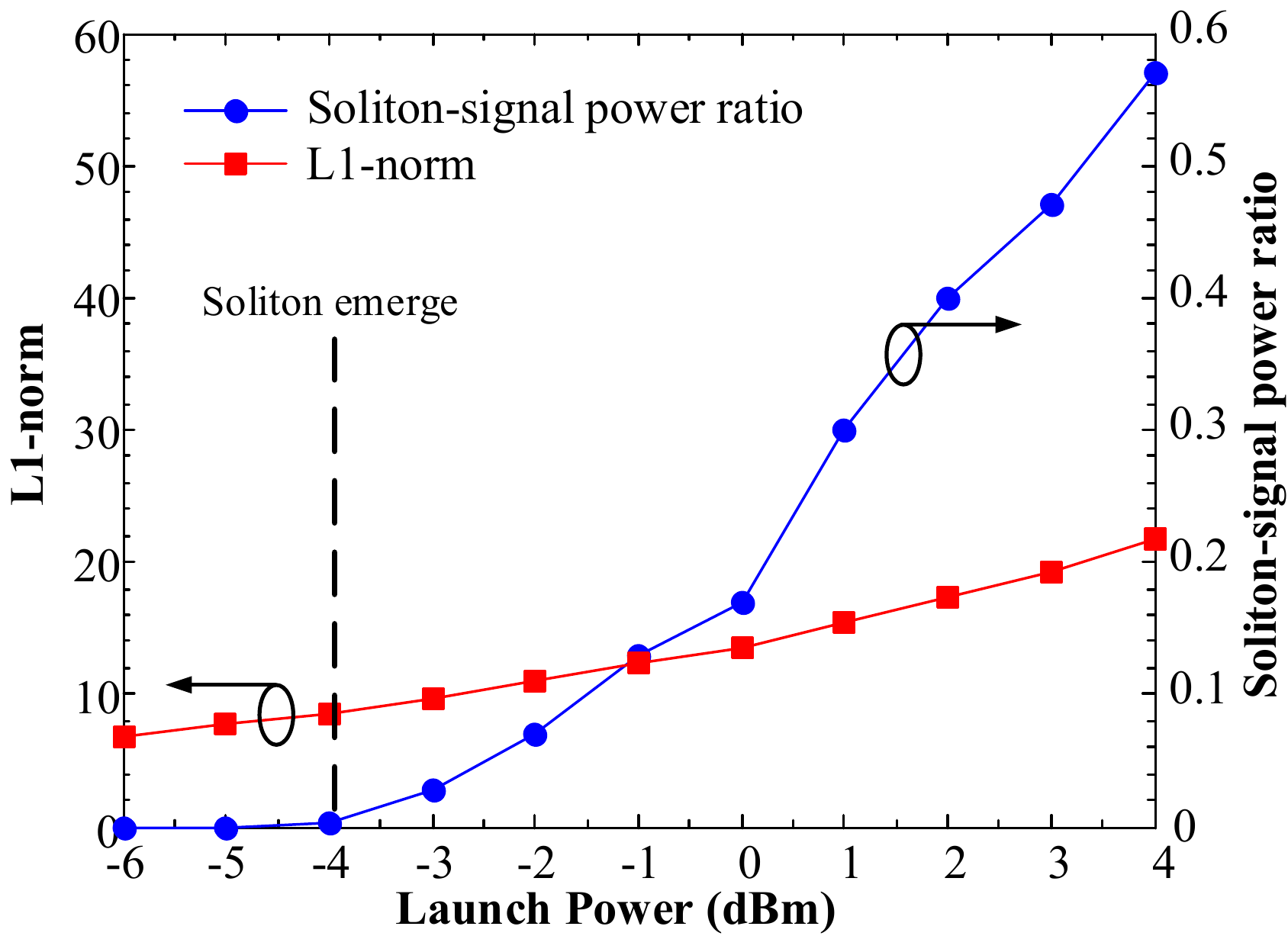}
\par\end{centering}

\caption{\label{fig:Simulation-results-4} $L_{1}$-norm and soliton-signal
power ratio for 100Gb/s OFDM.}
\end{figure}

The performances of both traditional DBP and DBP in the NFD are shown
in Fig. \ref{fig:Simulation-results-3}. DBP in the NFD can achieve
a performance gain of \textasciitilde{}$3.5$ dB, which is \textasciitilde{}$1$
dB better than DBP employing $4$ steps/span. The performance however
degrades dramatically once the launch power has become larger than
$-5$dBm. We attribute this phenomenon to the emergence of upper half-plane
roots of $\alpha(0,z)$, which corresponds to the formation of solitonic
components in the signal. This argument is supported in Fig. \ref{fig:Simulation-results-4},
where the $L_{1}$-norm and the ratio between the power in the solitonic
part ($\hat{E}$ in \cite[p. 4320]{Yousefi2014compact}) over the
total signal power are plotted as functions of the signal power. At
a launch power of $-4$ dBm, the signals begin to have solitonic components,
which seems to have a significant impact on the DBP algorithm in the
NFD. As anticipated in Sec. \ref{sub:Limitations}, solitons indeed
only occur above an $L_{1}$-norm which is significantly higher than
the bound $\pi/2$. 

\section{Conclusion\label{sec:Conclusion}}

The feasibility of performing digital backpropagation in the nonlinear
Fourier domain has been demonstrated with a new, fast algorithm. In
simulations, this new algorithm performed very close to ideal digital
backpropagation implemented with a conventional split-step Fourier
method for fibers with normal dispersion at a much lower computational
complexity. In the anomalous dispersion case, it was found that the
algorithm works well only if the signal power is low enough such that
solitonic components do not emerge. We are currently working to remove
this limitation.

\bibliographystyle{IEEEtran}
\bibliography{spawc15.bbl}

\end{document}